\documentclass{aa}
\usepackage{graphicx}
\usepackage{rotating}
\def\gtrsim{\mathrel{\hbox{\rlap{\hbox{\lower4pt\hbox{$\sim$}}}\hbox{$>$}}}}
\def\ltsim{\mathrel{\hbox{\rlap{\hbox{\lower4pt\hbox{$\sim$}}}\hbox{$<$}}}}

\begin{document}
\title{Variations of the ultraviolet resonance lines of the B2\,IV-V
star $\zeta$ Cassiopeiae}
\authorrunning {Smith \& Bohlender }
\titlerunning{UV Resonance lines of $\zeta$ Cas }
\author{M.A. Smith\inst{1} and D.A. Bohlender\inst{2} }
\institute{Department of Physics, Catholic University of America, 
Washington, DC 20064, USA; Present address: Space Telescope Science
Institute, 3700 San Martin Dr., Baltimore, MD 21218 ~~Email:~ msmith@stsci.edu \and
National Research Council of Canada, Herzberg Institute of Astrophysics, 5071 W. Saanich Rd., Victoria, BC Canada V9E 2E7 }

\date{Received ??; accepted ??}
\abstract{Recently Neiner et al. reported that the B2\,IV-V star $\zeta$\,Cas
contains a weak magnetic field which varies on the same 5.37 day period found
from the modulations of its N\,V, C\,IV, and Si\,IV UV resonance lines.
We have studied the time variable properties of the same resonance
lines in greater detail to determine the physical characteristics
of the magnetospheric structure responsible for them.
In our formulation
this structure takes the form of an axisymmetric ``disk" similar
to those around magnetic He-strong Bp stars. This structure corotates 
with the star, covering greater or lesser amounts of its
area during its transit.  
$\zeta$\,Cas offers a special case because
we observe it from a low inclination and yet its magnetic axis is
substantially inclined to the rotation axis.
The equivalent width-phase
curves show a flat maximum for half the cycle, indicating that the
disk is extended out of the plane, extends to the star's surface in
the magnetic plane, or both.  Synthetic spectra of the line profiles
during the maximum and minimum occultation phases can be
best reconciled with a disk geometry in which the resonance lines
are formed at a closed outer edge and along a thin outer layer.
We  speculate that observed weak redshifted emission is formed in
``auroral caps" located near the magnetic poles of the star.
We argue that this results from shocks of stagnated wind material
returning to the star and shocking against the outflowing wind.
}
\maketitle

\section{Introduction}

  Fully twenty years ago Grady et al. (1987) and Sonneborn et al. (1987) 
reported that the UV resonance lines of the sharp-lined B2\,IV-V star 
$\zeta$\,Cassiopieae 
(HD\,3360;$V$=3.67) are variable. Interest in this star has been rekindled
recently by the report by Neiner et al. (2003; hereafter N03) 
that this star has a dipolar magnetic field of 
B$_{z}$ $\approx$ 335\,G varying by ${\pm 28\,G}$ with
a period of 5.37 days. The 
resonance lines of N\,V, C\,IV, and Si\,IV show variations with the 
same period.  N03 also report a nitrogen abundance enhancement 
and a marginal helium overabundance ([He] = 0.11 ${\pm 0.06}$ dex). 
These characteristics make $\zeta$\,Cas a likely weak member of the 
magnetic He-strong Bp stars. 
We note for completeness that small-amplitude
 nonradial pulsations are excited on this semi-evolved star's surface 
(e.g., Sadsaoud et al. 1994). 

   It is generally accepted that magnetic Bp stars contain fossil
dipolar magnetic fields that are randomly oriented with respect 
to the rotational axes. Many authors (e.g., Shore 1987, Shore
\& Brown 1990, Babel \& Montmerle 1997a, 1997b) have attributed
the time-modulated UV resonance line variations to the channeling of
radiative winds from their magnetic poles to their equators. 
Because ionized particles cannot cross magnetic field lines, 
the global stellar fields play a key role in the wind outflows.
Although the particles emerging from the magnetic poles can freely 
accelerate, wind flows are prevented from crossing transverse stellar
field lines in the equatorial zones and are suppressed. Moreover, the 
field lines emanating from the stellar magnetic poles channel the
escaping wind toward the magnetic plane. Here they collide with wind 
streams coming from the opposite magnetic hemisphere. This produces a 
shock that overionizes particles which then reionize and deexcite and emit 
resonance line radiation. The resonance line radiation is then scattered
through the  confined structure, allowing the observer to see an 
array of absorption
and emission components as the corotating disk/shock complex corotates
around the star. Many of the cooled particles settle toward the plane to 
form a confined ``disk." Interior to the magnetospheric structure, a sector 
of this disk transits in front of the star and occults part of the star's
projected area and causes excess absorption in low-excitation lines of 
iron group ions. In Bp stars with extensive disks, the absorption from this 
``iron curtain" forest is made visible by their composite effect in 
high resolution spectra obtained by the {\it International Ultraviolet
Explorer (IUE)} (Smith \& Groote 2001; hereafter SG01) and in 
high-level Balmer lines (Groote \& Hunger 1976). 

  In the hydrodynamic formulation of this process by Babel
and coworkers, a shock develops where the disk particle pressure 
equals the wind's energy density.
Important refinements have been added to the Babel picture 
by ud-Doula and collaborators by considering the relative influences 
of other forces on particles along the flow. We summarize several 
effects they can have on the disk structure$\colon$

\begin{itemize} 
  \item Ud-Doula \& Owocki (2002; hereafter uO02) found that the confinement
  of the wind by the star's magnetic field is given 
  by a ratio ``$\eta_{*}$" of surface magnetic to wind energy densities. 
  These authors also
  found that the weight of the wind particles produces a feedback on 
  the field lines in which the stagnated wind drags the lines of 
  force inward along the equatorial plane. In some circumstances, the 
  particles may collapse the field lines back into the star's equatorial 
  regions. More typically, condensations form chaotically as snake-like 
  blobs. These structures fall back toward the star, out of the plane 
  and along the lines of force.  UO02 suggested that this sequence of 
  events is responsible for the UV emission lines in Bp star spectra. 
  Gagn\'e et al. (2005; G05) developed this idea to explain 
  emission from the hot Bp analog $\theta^1$\,Ori\,C.  A byproduct of this 
  picture is that matter does not fall straight inward along the magnetic
  plane to the star's surface. The effect is to produce a discrete inner 
  edge to the disk. Interior to this point there is effectively
  no matter.

\item
 Refinements in the picture have shown the importance
of angular momentum conservation and the tendency of particles 
to accumulate in magnetospheric regions of minimum potential energy 
(Preuss 2004, Townsend \& Owocki 2005). These forces can cause the disk
to warp for a range of magnetic inclinations and also permit 
the particles to accumulate
at the intersections of the magnetic and rotational equators and in 
other areas outside the planes. The 
effect of these accumulations is that disks of rapidly rotating Bp stars
may show departures from coplanarity and axisymmetry.

\item Ud-Doula et al. (2006; u06) found that 
lines of force near the Alfv\'en point of rotating 
stars are stretched outward by the centrifugal force. 
At irregular intervals, the field lines severe,
and disk material ``breaks out" through the Alfv\'en point at 
the equator. The heating of material by this process may be 
the source of X-ray flares, such as observed in
the prototypical magnetic Bp star
$\sigma$\,Ori\,E (e.g., Groote \& Schmitt 2004).

\item
  Including post-shock cooling energetics into quasi-equilibrium 
magnetic wind models 
leads to the prediction of a range of temperatures from
10$^{8}$\,K to 10$^{6}$\,K or less (Tonneson et al. 2002, G05). This 
shock region is confined to a broad zone just inside the Alfv\'en
radius. Interior to this zone, wind particles are not accelerated
to high enough velocities to produce ``superions" and resonance line 
radiation.

\end{itemize}

   For relatively slowly rotating, weakly magnetic stars like $\zeta$\,Cas
some of these processes are not expected to be important. An
important modification of the Babel model is that the wind 
detritus returns to the star along lines of force where it collides
with high latitude wind streams.

\section{Methodology}

\subsection{Reduction of ultraviolet resonance lines}

  The ultraviolet spectra for this program included the wavelength
regions surrounding the N\,V, C\,IV, Si\,IV, and Al\,III resonance 
lines and recorded on {\it IUE} echellograms with the Short Wavelength 
Prime (SWP) camera. These data were obtained from the MAST 
archives.\footnote{Multi-Mission Archive at Space Telescope
Science Institute, in contract to NASA.}
We found 106 spectra, all obtained between 1979 and 1995 and all observed
through the {\it IUE's} large aperture, were suitable for analysis.
We then computed ``equivalent width (EW) indices" by ratioing the fluxes 
within the line edges with fluxes of other wavelengths in the order (see
SG01). Spectra for phase ranges given in $\S$\ref{uvphs} were selected,
``conditioned," and coadded by carrying out the following 
operations$\colon$ (1) cross-correlating them to 
remove apparent wavelength shifts, (2) forcing them to 
a common continuum slope across the order, and (3) weighting the individual
spectra according to the mean fluctuation from pixels of neighboring
resolution elements. This weighting was accomplished by evaluating
the internal spectrum r.m.s. noise (assumed to be the median value of 
the absolute value of the fluctuations of pixels in each 
neighboring resolution element, i.e., every 3rd pixel).
The principal work in this paper concerns the analysis of pairs of
coadded resonance line spectra obtained at phases near the passage of 
the South magnetic pole across the star's central meridian (following 
N03, taken to be $\phi$ = 0.0), and approximately 0.5 cycles later, 
i.e., when these lines exhibited minimum and maximum absorption,
respectively.  
We quantified the variations of these line profiles in our first
figure by measuring
a dimensionless ``equivalent width index" rather than a true equivalent 
width (EW) in Angstrom units. Our index is defined as the ratio of all fluxes
within a designated window across the profile to all fluxes outside them
in the echelle order containing them. Our indices are proportional to
a true equivalent width but do not rely upon inaccurate placements of
the local continua. 
For the case of $\zeta$\,Cas, 
the windows were taken to be the instrumentally broadened photospheric line 
profile, or ${\pm 0.8}$ \AA~ for most resonance lines (${\pm 0.5}$ \AA\ for N\,V).

\subsection{Optical Data }

  In an attempt to search for variations in high-level Balmer lines
from the variable occultation of the star by the disk, we obtained 31
spectra with the 1.85 m telescope of the Dominion Astrophysical Observatory
during 2005, December 16, 2006 February 14--16 and March 8.  
We used the f/5 Cassegrain spectrograph and fed the light to a CCD 
detector by means of an image slicer. The resulting spectra have a 
dispersion of 10 \AA\,mm$^{-1}$, a resolution of about 16\,400, and
cover the wavelength region $\lambda\lambda$3610--3880. These data gave a 
null result, as reported below.

\subsection{Spectral Synthesis Computations}

  To perform the analysis of the {\it IUE} spectra, we utilized a suite  
of programs written by I. Hubeny and collaborators and non-LTE 
solar abundance BSTAR models (Lanz \& Hubeny 2006). 
The first of these, {\it synspec,} is a 
photospheric line synthesis program (Hubeny, Lanz, \& Jeffery 1994)
used for the reference photospheric profiles. 
This program permits one to modify assumed 
chemical abundances in the atmosphere. Indeed, the strengths of the
Fe\,III lines in the {\it IUE} spectrum suggested an iron 
abundance of ${\frac 13}$ the solar value, 
assuming a photospheric microturbulence
$\xi_t$ = 2 km\,s$^{-1}$. We also used {\it synspec} to convolve the
photospheric lines with functions approximating the instrumental and
rotational broadenings for the instrument and the star. 
The spectroscopic resolution of {\it IUE} high-dispersion spectra
is about 13\,000 and the the star's rotational broadening
($v\sin{i}$ = 17  km\,s$^{-1}$) are comparable .  
We computed the synthetic spectra in steps of 0.01\AA. 

  Our goal in this analysis is to assess the effects of the magnetospheric 
disk on the UV resonance line spectrum as the disk moves across the 
surface of $\zeta$\,Cas, alternately occulting greater or lesser areas
of the star in its circuit. To simulate the absorption and emission
signatures of the disk on the spectrum, we used the radiative transfer
program {\it circus} (Hubeny \& Heap 1996, Hubeny \& Lanz 1996). This 
program computes strengths of absorption or emission components of lines 
in a circumstellar medium from user-input quantities such as disk 
temperature, column density, areal coverage factor (the portion of 
the disk that passes in front of the star), and microturbulent velocity. 
{\it Circus} combines this contribution with the {\it synspec}-computed 
spectrum of the photosphere.
In its solution of the radiative transfer, {\it circus} calculates line 
emissions and absorptions separately along a line of sight. The emission
is computed according to the assumption of LTE for an input disk 
temperature $T_{disk}$. Because the ``absorption" of resonance lines
is actually due to scattering, it can be handled by turning off 
a flag in the program that computes the emissivity coefficient in the
radiative transport equation. The result is that the background flux 
of the star is affected only by the absorption coefficient.
The ability of handling the line transfer in this way permits one
to treat absorptions and emissions simultaneously or separately in
different circumstellar structures. 
{\it Circus} can accommodate three separate ``clouds." In the present
analysis, we found it necessary to include two cloud structures, one 
for the excess absorptions of the resonance lines and a second for
the emission components. The program allows us to input different
radial velocities and temperatures that might be useful in testing
whether different diagnostics are formed in separate regions,
allowing an assessment of whether they might arise in different volumes. 
As in the work described by SG01 and Smith et al. (2006; hereafter S06), 
we have assumed the clouds to be homogeneous in all properties. 
However, the MHD studies described above suggest this may well be 
an oversimplification.

  Among other input parameters are the disk's volume density and 
microturbulent velocity $\xi$. Our results were insensitive to the values 
chosen for either of them. The volume density enters only indirectly 
(and insensitively) through the gas ionization equilibrium. Our trial 
input density, $\sim 10^{12}$ cm$^{-3}$, is the order of magnitude SG01 
determined for hot Bp star disks and is also the peak value determined 
by the MHD simulations. Except for the absorptions of the primary 
resonance lines (C\,IV, Si\,IV) the overwhelming majority of the UV metal 
lines in our computations are optically thin. Since the spectral signatures 
then scale as the volume of absorbers, a degeneracy arises between two 
geometrical factors of the disk: its area and column length 
along the observer's line of sight. 

  The final ingredient of our models is the treatment of line of sight 
velocities.  Because the magnetosphere co-rotates over the star's surface, 
no differential velocities exist along the line of sight to the star
during occultation. Moreover, the Doppler shift of the disk absorbers 
is matched by the velocity of the star area behind them. SG01 found
that turbulent velocities in the range of 0\,--\,20 km\,s$^{-1}$ can
be offset by changes in the assumed metallicity in the disk because 
the lines are optically thin. These authors also found a broadening 
of the Al\,III and N\,V resonance lines of about 25 km\,s$^{-1}$ and
50 km\,s$^{-1}$, respectively, at phases near maximum occultation. The
{\it circus} models were computed with the same underabundant iron abundance 
as determined for the photosphere. We used solar abundances of 
C, Si, and Al, but an overabundance [N] = 0.4 dex (N03), all 
with a $\xi$ = 20 km\,s$^{-1}$.

\subsection{Relevant rotational and magnetic parameters }

  In our analysis we have assumed the physical parameters of 
$\zeta$\,Cas determined by N03. These include the redetermination of
the star's effective temperature, near 21\,000\,K and a 
logarithmic surface gravity of about 3.8. 
Therefore, we have chosen the (21,000, 4) model in the BSTAR grid as input 
to our {\it synspec} computations. At the surface of the star, small 
amplitude nonradial pulsations are present (N03), but we ignore them in this
discussion.  Given a period of
5.37 days and the star's $v_{\rm e}\,sin i$ = 17 km\,s$^{-1}$, the
inferred equatorial rotational velocity, $v_{\rm e}$ = 55 km\,s$^{-1}$, 
is also well determined$\colon$ $\zeta$\,Cas is a moderately rotating B star. 
N03 give a rotational obliquity angle of $i$ = 18 ${\pm 4}^{o}$. 
From $i$ and attributes of the magnetic phase, curve N03 derive a
magnetic obliquity angle, $\beta$ = 79 ${\pm 3}^{o}$. 

  The rotational phase zero $\phi$ is taken as the stellar 
longitude containing the negative poles in the visible hemisphere. 
Then the viewing angle $\alpha$ can be written as$\colon$

\begin{equation}
cos\,\alpha = sin\,\beta\,cos\,\phi\,sin\,i ~+~ cos\,\beta\,cos\,i,
\end{equation} 

\noindent one sees that for small $i$ and large $\beta$ that the viewing
angle changes little through the rotation cycle. 
Consequently, we can expect only small changes
in the absorption coverage factors of the disk in front of the star.

  The relation for the wind magnetic confinement parameter $\eta_{*}$
is given by uO05 as the ratio of magnetic to kinetic energies, 
namely as$\colon$

\begin{equation}
~~~~~~~~~~~~~~~~~~~~~~ \eta_{*} = B_z^2 R_*^{2}/(\dot{M} v_\infty)~.
\end{equation}

\noindent Here the symbols refer the the polar magnetic flux B$_z$,
stellar radius R$_*$, and mass loss rate $\dot{M}$, and the wind
velocity at infinity $v_{\infty}.$
One is somewhat handicapped in evaluating $\eta_{*}$ because no
modeling has been performed of $\zeta$\,Cas's wind attributes. 
Therefore, we will take mean values for a B2\,V star: 
$\dot{M}$ = 3$\times 10^{-10}$ 
$M_{\odot}$\,yr$^{-1}$ (Snow 1981). We estimate that $v_{\infty}$ = 800 
km\,s$^{-1}$ from the line profiles and $R_*$ = 5.9R$_{\odot}$ from N03. 
Then, $\eta_*$ $\sim$10$^{3}$. This is a large value but still
about 10$^{-4}$ times the value found for $\sigma$\,Ori\,E and 
36\,Lyn (S06)
However, the mass loss rate for $\zeta$\,Cas may be anomalously
high. For example, Slettebak (1994) found that the combined equivalent
widths for the the C\,IV and Si\,IV doublets for this star are 2.40\,\AA\
and 3.59\,\AA, respectively, whereas the corresponding values for an
average B2\,V star are 1.35\,\AA~ and 2.85\,\AA. If this is the case,
then $\eta_{*}$ should be taken as an upper limit. Finally, we can estimate 
the location of the Alfv\'en radius from the approximation $R_{Alf}$ 
$\approx$ $\eta$$^{\frac 14}$.  Thus, $R_{Alf, *}$ $\approx$
5$R_{*}$.

\section{Results}

\subsection{The UV resonance line variations with phase}
\label{uvphs}

\begin{figure}
\centering
  \includegraphics[width=6.5cm,angle=90]{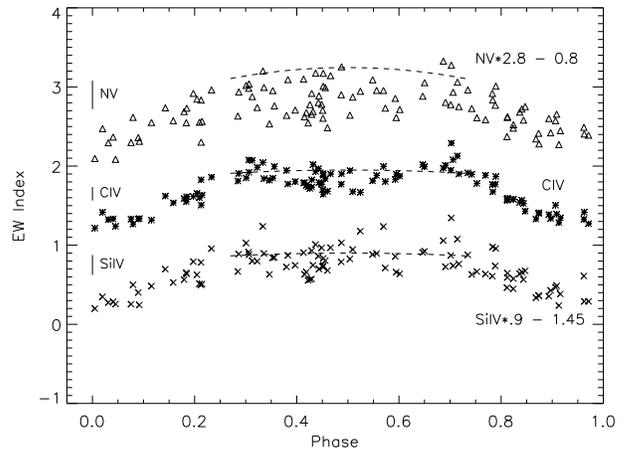}
\caption{   
Variation of line strength indices for the UV resonance doublets of 
N\,V, Si\,IV, and C\,IV with phase in the spectrum of $\zeta$\,Cas. 
Scaling factors are applied to show the same apparent amplitude; 
vertical shifts are made for clarity. Approximate measurement
errors are indicated. The dashed, tapered line shows the assumed 
limiting absorption relation (see text).
}
\label{ewphs}
\end{figure}

   N03 have determined a period of 3.57045${\pm 0.00008}$ days from
the variations of the C\,IV, Si\,IV, and N\,V resonance lines. 
To check this value, we applied the PDM period-finding algorithm
(Stellingwerf 1978) to EW indices extracted from the resonance lines 
as defined above.  
These extractions were done in the inner ${\pm 0.5}$ \AA\ 
of the N\,V lines and inner ${\pm 0.9}$ \AA\ of the C\,IV and 
Si\,IV lines. We ordered these values by time and and found a
value of  3.57038 ${\pm 0.0001}$ days, thereby verifying the N03 value.
We used this period and the N03 zero point ($\phi$ = 0 at HJD\,2446879.889) 
to compute rotational phases. Figure\,\ref{ewphs} exhibits the
absorptions (EW index) folded around this period, but we have rescaled
them so that each exhibits the same fractional variation (this rendition
artificially increases the apparent noise in the N\,V curve).
Note the Fig.\,\ref{ewphs} minimum centered at phase 0.0. 
Also, as previously pointed out by N03, 
one may discern a hint of a secondary minimum on the broader maximum 
at $\phi$$\approx$ 0.50,
but we are not convinced that it is statistically significant since
this feature does not appear in the Si\,IV or N\,V curves.
Finally, we point out that the shapes of the absorption-phase
curves are indistinguishable from one another. This is surprising
because, as noted below, the decreased absorptions they 
represent occur for both optically thick and thin lines.

   To reconnoiter the modulation of the line strengths with phase, we
first grouped over 20 spectra each into EW-minimum and EW-maximum phase 
bins corresponding to $\phi$=0.89--0.11 and 0.30--0.70, respectively,
and computed mean spectra from these groups. The results are exhibited 
as thick solid lines in Figures\,\ref{c4mxmn} and \ref{n5mxmn} for
the wavelength regions surrounding the C\,IV and N\,V doublets, 
respectively. Minimum and maximum spectra were also computed for the
Si\,IV lines.  The line profiles of the spectra are different in
that those for the former phases exhibit smaller absorptions and
generally steeper red wings. Part of the asymmetry in the wings is
due to the influence of the winds. 
As for other Bp stars and 
$\theta^1$\,Ori\,C (Smith \& Fullerton 2005), this wind is focused 
by the magnetic field and streams outward preferentially at 
low magnetic latitudes.
Since the wind escapes from
regions outside this disk, we do not model it in this work. 
The red shoulders of the profiles at these times 
are steep with respect to our computed 
photospheric profiles, and we attribute this to an incipient
redshifted emission. We attribute the strengthening
of the spectra to increased blocking of the stellar flux by the intervening 
disk.  Our modeling of absorptions and emissions proceeded with these 
assumptions, and we defer the discussion of results to the next section.

\subsection{Quantitative analysis of the C\,IV and N\,V resonance lines }

\subsubsection{The maximum and minimum absorption spectra}

  We have described a picture in which wind particles escaping the 
magnetic poles of a Bp star shock when they impact particles near the
magnetic equator that have formerly settled to form a cooled ``disk."
The phase-dependent absorptions of key ultraviolet 
lines validates this basic picture for $\zeta$\,Cas. The morphology
of the resonance lines shown is replicated within the errors by less
excited resonance lines of Al\,III and C\,II, thereby providing 
evidence that the structure is a confined cool disk.
However, we were not able to estimate the disk temperature from 
ratios of iron line absorptions because the available Fe\,II lines are 
too weak to show variations.  Even so, if we assume the relation for the
cool disk ionization temperature $T_{disk}$ $\approx$ 0.6\,$T_{\rm eff}$,
as found in the Bp stars studied by SG01 and S06, we obtain a value
$T_{disk}$ = 12\,000\,K.  Assuming this temperature and, in the absence 
of information about the extent of the coverage of the star by the
cool disk component during the occultation phase,
we find from {\it circus} modeling of the small variations of the 
Al\,III $\lambda$1855--62 and the C\,II $\lambda\lambda$1335-6 
multiplets that the hydrogen column densities of the
occulting disk are 1$\times10^{18}$ cm$^{-2}$, assuming $\xi$ = 20 
km\,s$^{-1}$, and 2$\times10^{18}$ cm$^{-2}$, assuming $\xi$ = 10
km\,s$^{-1}$. We take these as convenient lower and upper estimates. 

  We note that after reducing our DAO spectra of the high-level
(H8--H13) Balmer lines and plotting the equivalent widths with phase, we 
were unable to find statistically significant variations in them.
From our the measurement errors, ${\pm 0.6}$\% per line, this 
undetectability translates to a column density of $\sim4 \times 
10^{20}$\,cm$^{-2}$, according to our {\it circus} simulations, and
assuming a T$_{disk}$ = 12\,000\,K. Since this limit is a few hundred
times higher than the limits imposed by the All\,III and C\,II variations
just above, it does not add information to the disk analysis.

  The resonance line variations in the two figures bring out two 
interesting characteristics. First, notice that the r.m.s. spectra in 
the C\,IV line figure show that the profile activity is confined to 
roughly ${\pm 100}$ km\,s$^{-1}$ from line center. However, the 
wind-depressed blue wings undergo no activity through the cycle. 
Similar to the resonance lines of other Bp stars, the r.m.s. 
spectral amplitudes peak to the red of the line centers. 
Inspection shows that this is caused primarily by steepenings of the 
red shoulder of the lines in observations obtained in the EW-minimum 
phases. According to the line fittings discussed below, these shoulders
cannot be fit by photospheric spectra or disk absorption spectra.
Rather, we believe these are incipient emissions corresponding to the
more pronounced emissions observed in many Bp star spectra. 
Another characteristic brought out in these plots is that the doublet
ratios of these activities are 1.2 and $1.0{\pm 0.2}$ 
for the C\,IV and Si\,IV (not shown) lines, respectively. 
The ratio for the N\,V lines is 2.5${\pm 0.4}$. 
From these ratios we can see that the circumstellar media
in which the ``disk" component of C\,IV and Si\,IV are formed are 
optically thick while the N\,V lines optically thin, much like their 
photospheric components.

\subsubsection{Fitting of the resonance doublets }
\label{fits}

  In our line profile fitting we found that an important diagnostic 
is the line optical depths. Because the optical depths of our
models are typically within an order of magnitude of unity, we
were able to use them to estimate the optical depths of the disk
columns and ultimately
to set limits on the range of the disk temperature $T_{disk}$. 
These values are represented in the table discussed below.

\begin{figure}
\centering
   \includegraphics[width=6.5cm,angle=90]{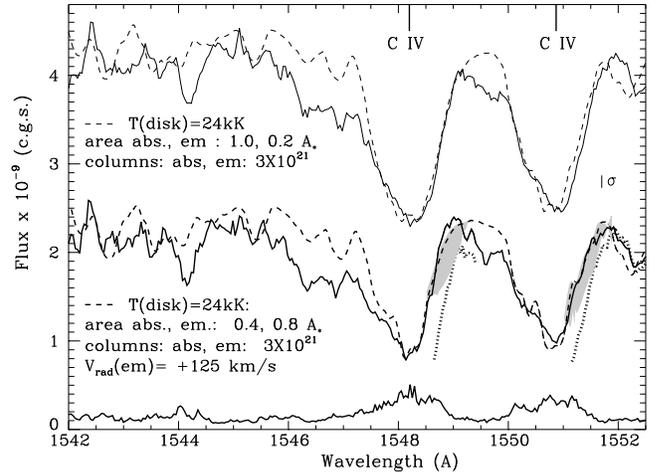}
\caption{   
Maximum (upper) and minimum absorption lines of the C\,IV doublet of
$\zeta$\,Cas fit with synthetic spectra (dashed lines).  
Fitting parameters are indicated (see Table\,1). 
The shaded area denotes the emission component determined in the models;
the dotted line in the lower plot repeats the maximum absorption
profile in this same wavelength range for comparison.
The blue wing absorption of the C\,IV lines is due to the
(unmodeled) wind contribution.
The r.m.s. spectra is also shown.
}
\label{c4mxmn}
\end{figure}

\begin{figure}
\centering
   \includegraphics[width=6.5cm,angle=90]{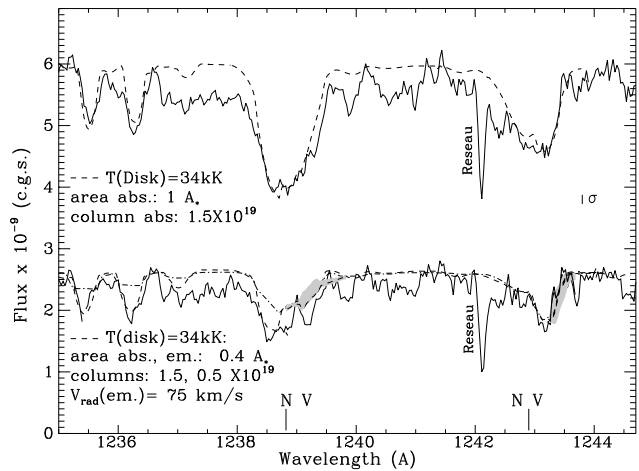}
\caption{   
Maximum (upper) and minimum absorption in the N\,V doublet lines of
$\zeta$\,Cas fit with synthetic spectra (dashed lines).
Fitting parameters are indicated (see also Table\,1).
Vestigial emission (shaded area) in $\lambda$1242 is suggested in our model.
}
\label{n5mxmn}
\end{figure}

   Our initial attempts to model the minimum and maximum EW spectra
in Figs.\,\ref{c4mxmn} and \ref{n5mxmn} used input {\it circus}
parameters for a single homogeneous absorbing and emitting circumstellar
``cloud." The models assume a disk temperature $T_{disk}$, which for 
simplicity we have assumed to be the same value. This group of models 
failed to give satisfactory fits. 
Although there were several problems with the fits, the most important 
was the inability of the models to produce enough absorption to 
fit the EW variations. This problem was accentuated by the constraint 
that the change of coverage of the star must be incomplete because of the 
low inclination of the star.  Investigating this failure 
in more detail, we found that the poor fits extended over a significant 
wavelength range.  For this reason, we were forced to consider {\it 
circus} models fit with two homogeneous clouds with different properties,
especially radial velocity. These models were successful, but only
when we increased the microturbulence $\xi$ to 75--90 km\,s$^{-1}$. 
We note by contrast that high cloud macroturbulences merely smear the computed 
excess absorptions and emissions. They do not strengthen 
them, as is required by the observations. 

   We were not led immediately to our best fits with the two-cloud 
description. Moreover, those we have selected for representation are
those obtained only for one possible disk temperature among the range
of  T$_{disk}$ values. One of the most important fitting
parameters, the temperature implied by fits to the 
emission components, T$_{emis}$, is not uniquely 
determined.  For the C\,IV line fitting, the
upper limit on $T_{emis}$, 26\,000\,K, was set by the decreased
optical depths in the line that led to different and contradictory 
absorbing cloud areas or columns for the two doublet components.
The lower limit of 16\,000\,K was set by the appearance of nearby Fe\,II
lines in the computed models, which are not present in the observations.
For the N\,V lines a low temperature limit of 26\,000\,K was again set 
by the absence of a predicted iron line (Fe\,III 1243.4\,\AA). 
The column density required for fitting 
the N\,V lines rapidly decreases to the values found for the C\,IV 
fittings, and we consider this the most likely result.

\begin{table*}[ht!]
\begin{center}
\caption{\label{tab1}\centerline{
Derived columns, areal coverages, and radial velocities (km\,s$^{-1}$) inferred from  UV resonance lines} 
}
\centerline{~}
\begin{tabular}{crcrcrlrc}  \hline\hline

N\,V lines  & ($\xi$ = 75 km/s)$\colon$ &  &  &  &  &   &  &  \\
$T_{disk}$  & Col. density &  & Fractional & Area    & Vel.  & Vel.    & $\tau_{line}$ & $\tau_{line}$   \\
 &    (Abs)~~~  &   (Emis)   &   (Abs) &  (Emis) &   (Abs) &  (Emis) &  (Abs) &  (Emis) \\
  34\,000  &  $1.5\times10^{19}$ &  ---   &  1.0 &  0.0  &    0  &   --  &   0.3 &   -- \\
  34\,000  &  $1.5\times10^{19}$ &  $5\times10^{18}$ &  0.4 &  0.4: &   0  &   +75 &    0.3 &   0.1 \\

  28\,000 & $3\times10^{21}$ &      --          & 1.0  &  0.0   &    0  &   -- & 0.6  & -- \\
  28\,000 & $3\times10^{21}$ & $3\times10^{21}$ & 0.4  &  0.4:   &   0  &   +50 & $0.6$  & $0.6$ \\

 &   &  &   &    &   &   &   &   \\

C\,IV lines & ($\xi$ = 90 km/s)$\colon$  &  &  &  &  &   &  &  \\
$T_{disk}$  & Col. density &  & Fractional & Area    & Vel.  & Vel.    & $\tau_{line}$ & $\tau_{line}$   \\
 &    (Abs)~~~  &   (Emis)   &   (Abs) &  (Emis) &   (Abs) &  (Emis) &  (Abs) &  (Emis) \\
  24\,000 & $3\times10^{21}$ & --   &   1.0 &  0.2 &   0 &  -- &  15 &   -- \\
  24\,000 & $3\times10^{21}$ & $3\times10^{21}$ &  0.4 &0.8 &   0 & +125&   5 &  5 \\

  20\,000 &  $3.5\times10^{19}$ &  -- & 1.0  &  0.0  &  0  &   --   &   40  &  -- \\
  20\,000 &  $1.7\times10^{17}$ & $1.7\times10^{17}$ & 0.4  &  1.0 &  0  &  +125  &    0.2 &  0.2 \\
 &   &  &   &    &   &   &   &   \\
 Si\,IV $\lambda$1403 & ($\xi$ = 75)$\colon$  &  &  &  &   &  &  \\
$T_{disk}$  & Col. density &  & Fractional & Area    & Vel.  & Vel.    & $\tau_{line}$ & $\tau_{line}$   \\
 &    (Abs)~~~  &   (Emis)   &   (Abs) &  (Emis) &   (Abs) &  (Emis) &  (Abs) &  (Emis) \\

  22\,000  &  $5\times10^{18}$ &  -- & 1.0 &  0.0  &    0  &   --  &   2 &   -- \\
  22\,000 &   $5\times10^{18}$ &  $5\times10^{18}$ & 1.0  &  2.0   &  +25 &  +50  &   2 &    2 \\

  18\,000 & $3\times10^{18}$ &  --    &  1.0 & 0.0  &  0 &   --  &   1.5 &  -- \\
  18\,000 & $3\times10^{18}$ & $3\times10^{18}$ &  1.0 &  6.5  &    +25 &  +50 &   1.5 & 1.5 \\
\hline
\end{tabular}
\end{center}
\end{table*}

  Table\,1 shows the results of best two-cloud models for the maximum 
and minimum EW profiles of the N\,V, C\,IV and Si\,IV doublets, that is 
for observations obtained near mid-transit (maximum occultation) phases
and those for half a cycle later, respectively. An example of a model for 
the C\,IV and N\,V doublets is shown in Figs.\,\ref{c4mxmn} and \ref{n5mxmn} 
as a dashed line. Annotations give pertinent model fitting parameters.
Shadings denote the modeled incipient emissions.
In the table we give both ``high" and ``low" $T_{disk}$ values, as 
discussed above. In each case we give modeling parameters for fits to 
the maximum EW and minimum EW line in the upper and lower lines of each 
double row entry. In the table we give several key fitting parameters, 
including the $T_{disk}$, column density, areal coverage, radial 
velocity, and central optical depth of the stronger doublet member. 
Errors were estimated by repeating our models with single parameters
changed by varying degrees and comparing the resulting differences
in the equivalent widths with 2$\sigma$ changes in the line profiles 
within the same wavelength window from the r.m.s. errors in the
{\it IUE} spectra. The latter was accomplishing using a Monte Carlo
approach to compute the fluctuations as described in Smith (2006).
These errors led to the following estimated 2$\sigma$ errors in our
table$\colon$ ${\pm 15}$ km\,s$^{-1}$ in microturbulence $\xi$, ${\pm 20\%}$ 
in fractional areal coverage for C\,IV and Si\,IV, and ${\pm 50\%}$ 
in the column densities.
The errors for the latter two quantities are ${\pm 100\%}$ for the 
N\,V doublet.
Inspection of this table brings out the following points$\colon$

\begin{itemize}
\item Unlike the cases of {\it equator-on} Bp stars with large
$\beta$ angles, occultation of some part of the stellar disk occurs 
to one extent or another at all rotational phases. This is expected 
for star-disk systems with a small $i.$ This circumstance also hints 
that the disk is either extended in height (perpendicular to the 
plane) or extends inward to reach the star in the magnetic plane.

\item The column density and hence volume ratio of shocked gas in
which the resonance line emission and scattering occurs is several
times that of the column density inferred from the less excited C\,II 
and Al\,III ions. This result contrasts with the SG01 finding for
other Bp stars, where the ratios were comparable.
This indicates that
the ratios of the volumes can be quite different among Bp stars and
opens the door to the possibility that they are not cospatial. 

\item  An important parameter in our fittings was the 
radial velocity of the emission component -- typically 
+50--125${\pm 20}$ km\,s$^{-1}$. Its errors were estimated by a 
comparison of values found for the doublet members.

\item The large $\xi$ values for Bp disk absorptions 
and emissions is typically high (SG01, S06), and we cannot verify
that it is a true microturbulence. Irrespective of data quality
limitations, there is an intrinsic interpretational ambiguity between
geometries that imply a large $\xi$ in a single structure and 
multi-columns each with smaller $\xi$ values and separated in 
radial velocity. Such a contrivance would both broaden and strengthen 
the excess absorptions and emissions. 

\end{itemize}

\section{Interpretation} 
\label{intrp}

  In S06 we described a computer program written to describe the varying 
excess aborptions of an optically thick resonance line as a co-rotating disk
shaped like a cylinder transits across the star. 
The line is assumed to be optically thick, the absorption
of flux from the obscured region of the star is total. Then, one may 
express the occulted area of the star in terms of the
excess absorption measured.

  One can manipulate input parameters representing the dimensions of
the disk to fit observed phase absorption curves. 
For the case of $\zeta$\,Cas, the curves in Fig.\,\ref{ewphs} 
show that the absorption curves are essentially {\it flat} out to ${\pm 
{\frac 14}}$ cycles around maximum occultation phase, $\phi$ $\approx$0.50. 
Unfortunately, flat absorption phase curves give information only on the 
limits of ratios of dimensions. \footnote{Although we have 
discounted the statistical significance of a secondary dip in the phase 
curves, if they are real it could be interpreted as a {\it dearth} of 
overionized particles like C$^{3+}$ in zones over the rotational poles.
This would agree with uO02's expectations that the particles exert a 
feedback on the magnetic field and drag them away from 
zones of low angular momentum. Evidence for this avoidance would be 
visible only for rotating magnetic stars observed at low inclinations.} 
This problem is aggravated by the low S/N of {\it IUE} data, which 
makes it difficult to map the progress of the disk viewed at a low $i$.
However, we may still investigate the general character of the disk
cylinder shape if we estimate a minimum taper in the curves (5\% of 
the maximum excess absorption out to $\phi$= 0.25 and 0.75) that might 
be hidden in the noise of the absorption curves. This taper is shown
in Fig.\,\ref{ewphs} as a dashed line.

   Applying this criterion and the $i$ and $\beta$ values
for $\zeta$\,Cas, we adopt an outer disk radius $r_{outer}$ = 
R$ _{Alf},$ and we can ask, what is the interdependence between the height
(measured parallel to the rotational axis) and the thickness
of the disk in the radial direction,  $r_{outer}$ -  $r_{inner}$?
According to our models and our taper criterion, a disk extending in
to the star's surface along the magnetic plane can protrude no further than 
3.2$R_*$ from the central plane.
For a second case for which the disk has an inner radius of 
3$R_*$ (i.e., an evacuation of the inner 50\% of the disk), our models produce 
a similarly tapered phase if the disk extends $\ge$3.0R$_*$ out
of the magnetic plane. For the cylindrical model we have 
adopted there is a degeneracy between these dimensions -- we have 
little leverage in determining where an inner disk edge lies. 
More importantly than this, the implication from the cylindrical 
model representation is that the disk extends to surprisingly
large distances from the central plane. There is no indication that
disks of other Bp stars extend out of the plane to similar distances.
It seems more likely that our assumed disk geometry is in conflict
with fitting difficulties that are aggravated for star-disk systems
observed from a low inclination.

 Although the concern just raised over our disk geometry arises from the 
combination of high $\beta$ and low $i$ for the $\zeta$\,Cas case, the
resonance line absorption curves of stars having high $\beta$ and $i$ 
values pose their own interpretational dilemmas. The most significant of
these is a fact noted by SG01 for those systems for which the observer's 
observer's line of sight runs through the magnetic plane. 
In these cases, the complete occultation of the star disk in optically 
thick resonance lines presents the physical problem of how superions 
responsible for the line scattering can be distributed throughout the cylinder. 
How can such ions be excited in the middle of an optically thick medium? 
This puzzle can be solved if, as predicted in the MHD models of 
ud-Doula and coworkers, the disk is pinched off at the Alfv\'en 
radius in the equator by the confluence of collimated wind streams.
Exterior to this point the field lines curve out and run parallel to the
magnetic equator. Then, an external observer at infinity viewing
the system in the magnetic plane sees the emission 
from excited superions along this (thin) geometric boundary. 

  A third difficulty offered for the cylindrical geometry, again evident
for the $\zeta$\,Cas case, is that the absorptions should scale differently
for optically thick and thin lines as the disk moves off the limb of
the star. The excess absorptions of these two groups of lines
should scale differently -- as the area and the volume of the occulting
disk, respectively. In contrast, we find that when the curves in 
Fig.\,\ref{ewphs} are overplotted, their morphology is indistinguishable
for the two types of lines. This difficulty is inherent for any ``solid"
disk geometry. It can be avoided if the disk opacity is confined to 
a thin outer region having similar depth ranges of formation for these 
lines. This circumstance is permitted if the volumes of excited ions are
confined to the outer regions of the disk.

\begin{figure*}
\centering
   \includegraphics[height=17.3cm,width=6.0cm,angle=-90]{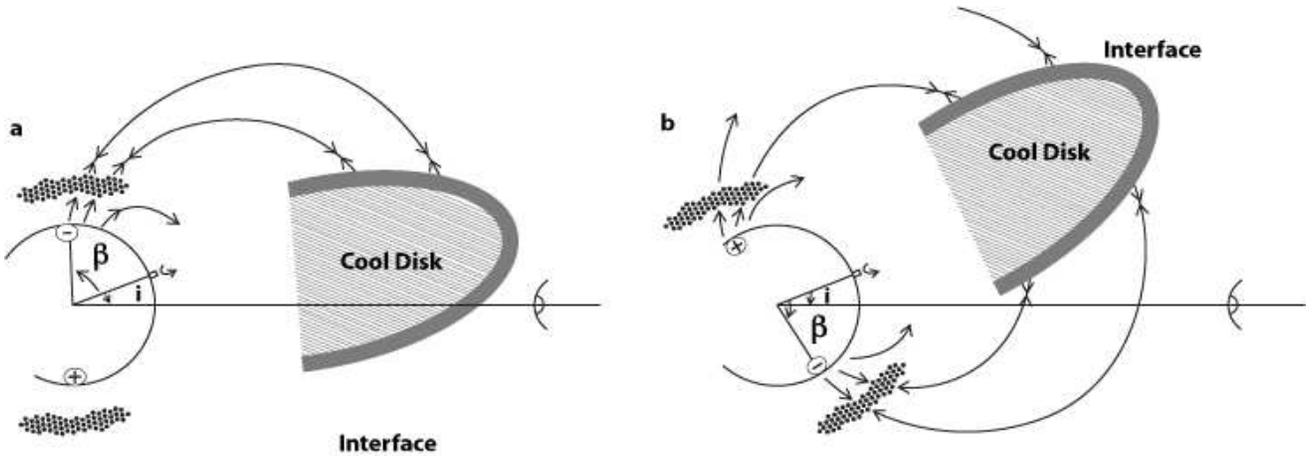}
\caption{   
The geometry at the times of 
maximum ($\phi$ = 0.5) and minimum ($\phi$ = 0) absorptions of the 
resonance lines of $\zeta$\,Cas in panels a) and b), respectively. The 
observer is to the right. The angles $i$ = 18$^o$ and 
$\beta$ = 79$^o$ are taken from N03. Low-excitation line absorptions are
formed in the optically-thin disk interior.  Closed at the outer end, the
disk is centered around the magnetic equatorial plane and taken to
be an oblate spheroid. The outer edge of the disk is heated by impacts
from the high velocity wind and includes superions responsible for the
resonance line absorptions.  Net redshifted emissions are produced in an 
``auroral cap" shown as a dotted region over the magnetic poles (at $\sim$11
and $\sim$5 o'clock positions).
}
\label{skch}
\end{figure*}

  In Figure\,\ref{skch} we suggest a revised geometry in which the outer disk 
edge, representing the wind-disk shock, tapers toward the equator and thus 
takes the general shape of an oblate spheroid.\footnote{Rather than an
oblate spheroid, the circumstellar disk region could take on just as well a 
negative curvature, coming to a cusp at the equator. The important point in 
our picture is not the exact shape but that it be closed at the equator and
have about the height and radii we determine.} 
We envisage that the excited ``super-ions" are formed in a thin outer
shell of this structure. The column and volume densities can then
be made consistent if the absorption column lengths, as 
in Table\,1, refer to a common shell thickness. For densities 
predicted by uO02 and column lengths in Table\,1, the shell thickness 
to $\tau$ = 1 is then only of the order 10$^3$ km.
Since the resonance lines are formed in a skin around a cool transparent
disk, there need be no relation between their column densities and those
inferred from the Al\,III and C\,II lines.

  The regions where the resonance line emissions arise
likewise should be reconsidered, as they cannot be understood as
the result of the interaction of the wind and the edge of a dense disk.
Consider first that for the Bp stars with dense, extensive disks 
investigated by SG01, the emissions can be present at all phases.
The SG01 picture was that the resonance lines 
exhibit emission preferentially at those phases when
the disk is viewed nearly face-on in the plane of the sky.  A problem
with this geometry is the positive radial velocity of the emissions. 
In contrast to SG01, we now believe the redshifts do not arise from the
impact of the wind particles against the disk interface. Such
interactions would produce emissions, but they are centered at the 
rest frame of the star. 

   A second problem is that the {\it ratios} of volumes computed 
from these lines' emissions and excess absorptions can 
be quite diverse among Bp stars. However, they should be approximately 
equal if the contributions in the same disk volumes arise from scatterings. 
These scatterings are conserved when integrated over all lines of sight.

  These issues can be addressed by positing that the
absorptions and most of the emissions of UV resonance lines 
are formed in different places.  Our revised geometrical picture 
harkens back to the results of two-dimensional
MHD models of the emission complex surrounding $\theta^1$\,Ori\,C 
discussed by uO02 and G05. Recall that these models predict that infall 
condensations fall back into the outflowing wind. Indeed, the collision
of these two structures will produce redshifted emission in the external
reference frame because the center of mass is dominated by the
falling blobs. These emissions are visible mainly over a narrow range of 
phases because the projected shock areas modulate with viewing angle.
For $\theta^1$\,Ori\,C the projected radial velocity in 
the UV and X-ray resonance line is  observed to be $\gtrsim$100 km\,s$^{-1}$ 
(Smith \& Fullerton 2005, G05). 

   We now ask what would be outcome of this same process if the wind were 
weaker? We speculate that the answer begins with the fact
that because the lower efflux would 
provide less resistance to the infall, the shock front where the pressures
equalize will lie closer to the star. This shock point will then occur 
where the dense fallback blobs have accelerated to a higher velocity, 
while the wind outflow, intercepting them from the opposite direction,  
has only begun to accelerate.  For a sufficiently weak wind (and similar 
$\eta_{*}$) the shock front should then describe an arc-like ``auroral" 
cap not far above the star's magnetic pole with a center-of-mass redshift 
because of the fall back's higher velocity and density. We
expect that there are many such returning blobs at any one time, so
observations repeated at the same phase will see the same level of
emission.  The projected area of the shock viewed
by the observer, and hence the recorded emission of the resonance lines, 
would become maximum at $\phi$ = 0. (The reverse would be the case 
at $\phi$ = 0.5 when the project area of the cap would be minimal.) 
Clearly, this picture begs to be confirmed by MHD simulations of
the type conducted for other Bp stars having stronger fields.

  The  putative auroral shock regions are represented as shaded 
regions at the ``11 o'clock" 
and ``5 o'clock" positions relative to the star in Fig.\ref{skch}. 
In this new configuration at $\phi$ = 0 these shocks have smaller
areas but are denser than the 
the wind-shock interfaces discussed by Babel and SG01. 
Note that because the emissions and absorptions are formed mainly in
two different regions of the magnetosphere, their volumes and
column densities need not be equal.  In addition, one expects that the
UV line emission is formed in a separate region of the magnetosphere
from the X-ray emissions. The energies released at the auroral caps
can be expected to be moderate at most.

 This picture suggests that sometimes one might see {\it small}
amounts of blueshifted emission from the opposite hemisphere.
As shown by Fig\,\ref{skch}, right panel, ``4--5 o'clock" auroral cap
contribution from the aurora over the opposing figure would exhibit 
a much smaller Doppler shift and projected area than the opposing cap.
We can speculate that the very high microturbulence taken for conditions 
in a single structure in our previous models, are the result of a 
superposition in velocity of these two shocks (each with a smaller $\xi$).

\section{Summary}

   Our modeling of the resonance line variations of $\zeta$\,Cas shows
that the disk occults the star to some degree during the 
entire rotation cycle. This attribute has not been found in 
the quantitative modeling of other magnetic Bp stars whose lines we
have examined (e.g., SG01, Smith 2003, Smith \& Fullerton 2005, S06).
It is probably more easily found for $\zeta$\,Cas because of its
small inclination angle $i.$ In this geometry the disk occultation
never realizes the full possible range of areal coverages (0--100\%).

  As mentioned, the large areal coverage factor found for
our solutions at both $\phi$ = 0.0 and 0.50 requires
that the disk either extend nearly to the star's 
surface, extend to large distances from the central magnetic plane, 
or both. In Fig.\,\ref{skch}, we noted that most or all the lines of 
sight to the star intersect the disk and thus suffer flux attenuation.

  Consideration of the characteristics of the emission components in
$\zeta$\,Cas and the ``classical" Bp stars shows that they cannot be
easily understood with the kinematic description adopted by SG01, whether 
the emission arises from a shock at the edge of a cylindical or torus
shaped disk.  For example, for the Bp stars with dense or extensive 
disks investigated by SG01, the emissions can be present at all phases,
and in all cases with a net redshift. We conjectured that this might
be understood by shocks created by the collisions of dense fall 
back (blobs) with the outflowing wind. This releases the magnetic disk 
model from other interpretational inconsistencies. 
These include the  frequent inequalities between the derived
volumes in which the absorptions and emission are formed. It also addresses
the high microturbulent velocities that are otherwise necessary to
fit the line profile variations.

  The analysis of lines of a modestly inclined system like $\zeta$\,Cas 
clarifies these problems and provides cause for a reconsideration 
of the disk and emission region geometries. This shows that a final
picture of these complicated star-disk system requires
observations not merely over all viewing angles but of
a sample of stars with a variety of rotational and magnetic 
inclination angles.

We thank the referee, Dr. Marc Gagn\'e, for a number of comments 
that have added to the quality of this paper.


\begin{thebibliography}{}

\bibitem[]{} Babel, J., \& Montmerle, T. 1997a, A \& A, 323, 121

\bibitem[]{} Babel, J., \& Montmerle, T. 1997b, A \& A, 485, L29

\bibitem[]{} Gagn\'e, M., Oksala, M. E., Cohen, D. H.
 et al. 2005, ApJ, 628, 986 (G05)

\bibitem[]{}  Grady, C. A., Bjorkman, K. S., and Snow, T. P. 1987, 
ApJ, 320, 376

\bibitem[]{} Groote, D., \& Hunger, K. 1976, A. \& A., 52, 303

\bibitem[]{} Groote, D., \& Schmitt, J. H. M. M. 2004, A. \& A., 418, 235

\bibitem[]{} Neiner, C., Geers, V. C., Henrichs, H. F., et al. 2003, A \& A, 411, 565 (N03)

\bibitem[]{} Owocki, S, P. \& Townsend, R. H. D. 2005, MNRAS, 357, 251 

\bibitem[]{} Preuss, O, Schussler, M, Holzwarth, V. et al. 2004, A. \& A.,
417, 987

\bibitem[]{}  Sadsaoud, H., Le Contel, J. M., Chapellier, E. 
et al. 1994, A. \& A., 
287, 509 

\bibitem[]{} Shore, S. A., 1987, AJ, 94, 731

\bibitem[]{} Shore, S. A., \& Brown, D. N. 1990, ApJ, 365, 665 

\bibitem[]{} Slettebak, A. 1994, ApJS, 94, 163

\bibitem[]{} Smith, M. A. 2003, International Confrence on Magnetic 
Fields in O, B and A Stars, ed.  L. Balona, H. Henrichs, \& R. Medupe, 
ASP Conf. Ser., 305, 310

\bibitem[]{} Smith, M. A. 2006, A. \& A., 459, 215

\bibitem[]{} Smith, M. A., \& Groote, D. 2001, A. \& A., 372, 208 (SG01)

\bibitem[]{} Smith, M. A., Wade, G., Bohlender, D. A., et al. 2006, A. \& A., in press (S06)

\bibitem[]{} Snow, T. P., Jr. 1981, ApJ, 251, 139

\bibitem[]{} Sonneborn, G., Garhart, M., \& Grady, C. 1987, in
Physics of Be Stars, ed. A. Slettebak \& T. P. Snow., IAU Colloq. No. 92,
286 (Cambridge Univ. Press)

\bibitem[]{} Stellingwerf, R. F. 1978, ApJ, 224, 953

\bibitem[]{} Tonneson, S., K, Cohen, D. H., Owocki, S. P., et al. 
2002, BAAS, 201, \#113.01

\bibitem[]{} Townsend, R. H. D. \& Owocki, S. P. 2005, MNRAS, 357, 251 

\bibitem[]{} ud-Doula, A., \& Owocki, S. P. 2002, ApJ, 576, 413 (uO02)

\bibitem[]{} ud-Doula, A., Townsend, R. H. D., \& Owocki, S. P. 2006, ApJ, 640, L191 (u06)

\end{thebibliography}
\end{document}